\documentstyle[proc,epsf,amssymb,amsmath]{rs}

\newcommand{\half}{\mbox{$\textstyle \frac{1}{2}$}}
\newcommand{\octa}{\mbox{$\textstyle \frac{1}{8}$}}

\newtheorem{prop}{Proposition}  

\def\thedemobiblio#1{\smallskip\par
 \list{}{\labelwidth 0pt \leftmargin 1em \itemindent -1em \itemsep 1pt}
 \small \parindent 0pt
 \parskip 1.5pt plus .1pt\relax
 \def\newblock{\hskip .11em plus .33em minus .07em}
 \sloppy\clubpenalty4000\widowpenalty4000
 \sfcode`\.=1000\relax}

\begin{document} 

\title[Interest Rates and Information Geometry]
{Interest Rates and Information Geometry} 

\author[D.C. Brody and L.P. Hughston]{
Dorje C. Brody$^{*}$ 
and 
Lane P. Hughston$^{\dagger}$  
} 
\affiliation{$*$Blackett Laboratory, Imperial College, 
London SW7 2BZ, UK \\ 
and Centre for Mathematical Science, Cambridge University, \\ 
Wilberforce Road, Cambridge CB3 0WA, UK \\ 
$\dagger$Department of Mathematics, King's College 
London, \\ The Strand, London WC2R 2LS, UK}

\date{\today} 
\maketitle 
\input{psfig.sty}  

\begin{abstract}  
The space of probability distributions on a given sample 
space possesses natural geometric properties. 
For example, in the case of a smooth parametric family of 
probability distributions on the real line, the parameter 
space has a Riemannian structure induced by the embedding 
of the family into the Hilbert space of square-integrable 
functions, and is characterised by the 
Fisher-Rao metric. In the nonparametric case the relevant 
geometry is determined by the spherical distance function 
of Bhattacharyya. In the context of term structure 
modelling, we show that minus the derivative of the 
discount function with respect to the maturity date gives 
rise to a probability density. This follows as a 
consequence of the positivity of interest rates. 
Therefore, by mapping the density functions associated 
with a given family of term structures to Hilbert space, 
the resulting 
metrical geometry can be used to analyse the relationship 
of yield curves to one another. We show that the general 
arbitrage-free yield curve dynamics can be represented 
as a process taking values in the convex space of smooth 
density functions on the positive real line. It follows 
that the theory of interest rate dynamics can 
be represented by a class of processes in Hilbert space. 
We also derive the dynamics for the central moments 
associated with the distribution determined by the 
yield curve. (26 June 2000) \\ 
\vskip0.1cm 
\begin{center} 
{\footnotesize 
{\bf Keywords: Interest rate models, Heath-Jarrow-Morton 
framework, \\ principal moments analysis, differential 
geometry and statistics } }
\end{center}
\end{abstract}

\section{Introduction} 

The theory of interest rates has gone through two major 
developments in recent decades. Following initial 
investigations by Merton (1973) and others, the first decisive 
advance 
culminated in the work of Vasicek (1977) who was able to give 
a fairly general characterisation of the arbitrage-free dynamics 
of a family of discount bonds, indexed by their maturity. The 
well-known model that bears his name appears as an exact 
solution obtained with 
specialising assumptions. In the wake of Vasicek's work were 
a number of other specific interest rate models, of varying 
degrees of usefulness and tractability, including, for example, 
the CIR model (Cox {\it et al.} 1985) and its generalisations. 
The next significant line of development, following the 
general martingale 
characterisation of arbitrage-free asset pricing by Harrison 
$\&$ Kreps (1979) and Harrison $\&$ Pliska (1981), was instigated 
with the recognition by Ho $\&$ Lee (1986) that the initial term 
structure 
might be specified essentially arbitrarily, a feature that has 
important practical implications. This insight was 
incorporated into the HJM framework (Heath 
{\it et al.} 1992), which constituted a major advance in the 
subject, providing a general model-independent basis for 
the analysis of interest rate dynamics and the pricing of 
interest rate derivatives. 

Since then there have been numerous further developments. These 
include, for example, the infinite dimensional or `string-type' 
models of Kennedy (1994), Santa-Clara $\&$ Sornett (1997) and 
others, the positive interest rate models of Flesaker $\&$ 
Hughston (1996), the potential approach of Rogers (1997), 
the so-called market models (Brace {\it et al}. 1996, 
1997; Jamshidian 1997), and the geometric analysis of the space 
of yield curves undertaken by Bj\"ork $\&$ Svensson (1999). 

Nevertheless, no criterion has emerged, based on the 
extensive econometric evidence available, that allows in 
a rational way for the identification of a clearly 
preferred class of 
models. On these grounds it makes sense to try to cast the 
general interest rate framework into a new form, with the 
idea that certain models might thus become recognisable 
as more natural on mathematical and economic grounds. 

With this end in mind, the 
purpose of the present article is to propose a 
novel application of information geometry to interest rate 
theory. The main results are (i) 
the construction of a geometric measure for how `different' two 
term structures are from one another; (ii) a characterisation of 
the evolutionary trajectory of the term structure as a 
measure-valued process; (iii) the derivation of dynamics 
for the principal moments of the 
term structure; and (iv) a reformulation 
of arbitrage-free interest rate dynamics in terms 
of a class of processes on Hilbert space. 

The paper is organised as follows. In \S 2 we review the 
basic idea 
of information geometry and its role 
in estimation theory. The geometry of the normal 
distribution is considered in detail as an illustration. In \S 3 
a remarkable characterisation of the discount function in terms 
of an abstract probability density function is introduced in 
Proposition 1. This allows us to apply information geometric 
techniques to determine the deviation between different term 
structures within a given model. In this connection, in \S 4 
we consider a class of flat rate models as examples. 

The material of the first four sections of the paper is 
essentially static, i.e., set in the present, whereas in \S 5 
we investigate the dynamics of 
the density function that generates the term structure. This is 
carried out in such a way that the resulting dynamics is 
manifestly arbitrage-free. Our key result here is formula 
(\ref{eq:5.15}), in which we establish that the dynamics of the 
term structure can be characterised as a measure-valued process. 
This idea is developed further in Proposition 2. 

In \S 6 we introduce an analogue of the classical principal 
components analysis for yield curves, and in Propositions 3 and 
4 we derive formulae for the evolution of the first two moments 
of the term structure density process. Then, making use of the 
information geometry developed earlier, in \S 7 we map the 
dynamics developed in \S 5 to Hilbert space. Our main result 
here is Proposition 5, which shows how this can be achieved. 

\section{Information geometry} 

Because some of the mathematical 
techniques we employ here may not be 
familiar to those working in finance, it will be appropriate to 
begin with a few background remarks. It has long been known 
(see, e.g., Amari 1985; Kass 1989; Murray $\&$ Rice 1993) that 
a useful 
approach to statistical inference is to regard a parametric model 
as a differentiable manifold equipped with a metric. The 
recognition that a parametric family of probability distributions 
has a natural geometry associated with it arose in the work of 
Mahalanobis (1936), Bhattacharyya (1943) and Rao (1945) over 
half of a century ago. 

Suppose, for example, that $X$ is a continuous random variable 
taking values on the real line ${\bf R}^{1}$, and that $\rho(x)$ is 
a density function for $X$. Because $\rho(x)$ is nonnegative and has 
integral unity, it follows that the square-root 
likelihood function 
\begin{eqnarray} 
\xi(x) = \sqrt{\rho(x)}
\label{eq:2.1}
\end{eqnarray} 
exists for all $x$, and satisfies the normalisation condition 
\begin{eqnarray}
\int_{-\infty}^{\infty} (\xi(x))^{2} {\rm d}x = 1 . 
\label{eq:2.2} 
\end{eqnarray}
We see that $\xi(x)$ can be regarded 
as a unit vector in the Hilbert space 
${\cal H}=L^{2}({\bf R}^{1})$. Now let 
$\rho_{1}(x), \rho_{2}(x)$ denote a 
pair of density functions on ${\bf R}^{1}$, and $\xi_{1}(x), 
\xi_{2}(x)$ the corresponding Hilbert space elements. 
Then the inner product 
\begin{eqnarray}
\cos\phi = \int_{-\infty}^{\infty}\xi_{1}(x)\xi_{2}(x) {\rm d}x 
\label{eq:2.3}
\end{eqnarray}
defines an angle $\phi$ which can be interpreted as the {\sl 
distance} between the two probability distributions. More 
precisely, if we write ${\cal S}$ for the unit sphere in 
${\cal H}$, then $\phi$ is the spherical distance 
between the points on ${\cal S}$ determined by the vectors 
$\xi_{1}(x)$ and $\xi_{2}(x)$. 
The maximum possible distance, corresponding to nonoverlapping 
densities, is given by $\phi=\pi/2$. This follows from the fact 
that $\xi_{1}(x)$ and $\xi_{2}(x)$ are nonnegative functions, 
and thus define points on the positive orthant of ${\cal S}$. 
We remark that an alternative way of expressing (\ref{eq:2.3}) 
is 
\begin{eqnarray}
\cos\phi = 1 - \frac{1}{2} \int_{-\infty}^{\infty} 
(\left( \xi_{1}(x)-\xi_{2}(x)\right)^{2}{\rm d}x , 
\label{eq:2.4}
\end{eqnarray}
which makes it apparent that the angle $\phi$ measures 
the extent to which the two distributions are distinct. 

The spherical distance of Bhattacharyya introduced above is 
applicable in a nonparametric context. In the case of a 
parametric family of probability distributions we can develop 
matters further. Let us write 
$\rho(x,\theta)$ for the parameterised density function. Here 
$\theta$ stands for a set of parameters $\theta^{i}$ 
$(i=1,\cdots,r)$. By varying $\theta$ we obtain an 
$r$-dimensional submanifold ${\cal M}$ in ${\cal S}$ 
determined by the unit vectors $\xi(x,\theta)\in{\cal H}$. 
The parameters $\theta^{i}$ are local coordinates 
for ${\cal M}$. 

The key point that we require in the 
following (cf. Dawid 1977) is that the spherical geometry 
of ${\cal S}$ induces a Riemannian geometry on 
${\cal M}$, for which the metric tensor $g_{ij}(\theta)$ is 
given, in local coordinates, by 
\begin{eqnarray}
g_{ij}(\theta) = \int_{-\infty}^{\infty} \frac{\partial\xi(x,\theta)}
{\partial\theta^{i}}\frac{\partial\xi(x,\theta)}
{\partial\theta^{j}} {\rm d}x . 
\label{eq:2.5}
\end{eqnarray}
By use of definition (\ref{eq:2.1}), we see that an alternative 
expression for $g_{ij}(\theta)$ is 
\begin{eqnarray}
g_{ij}(\theta) = \frac{1}{4}\int_{-\infty}^{\infty} \rho(x,\theta) 
\frac{\partial\ln\rho(x,\theta)}
{\partial\theta^{i}}\frac{\partial\ln\rho(x,\theta)}
{\partial\theta^{j}} {\rm d}x ,  
\label{eq:2.6}
\end{eqnarray}
which shows (cf. Brody $\&$ Hughston 1998) that the metric 
$g_{ij}$ is, apart from the factor of $\frac{1}{4}$, the Fisher 
information matrix, i.e., the covariance matrix of the parametric 
gradient of the log-likelihood function (Fisher 1921). We refer 
to $g_{ij}(\theta)$ as the Fisher-Rao metric on the 
statistical model ${\cal M}$. 

The significance of the Fisher-Rao metric in estimation theory 
is well known. Suppose that $\tau(\theta)$ is some given function 
of the parameters, and that the random variable $T$ represented 
by the function $T(x)$ on ${\bf R}^{1}$ is 
an unbiased estimator for $\tau(\theta)$ in the sense that 
\begin{eqnarray}
\int_{-\infty}^{\infty} \rho(x,\theta) T(x) {\rm d}x = 
\tau(\theta) . 
\label{eq:2.7} 
\end{eqnarray}
The variance of the estimator $T$ is defined, as usual, by 
\begin{eqnarray} 
{\rm Var}[T] = \int_{-\infty}^{\infty} \rho(x,\theta) 
(T(x)-\tau(\theta))^{2}{\rm d}x . 
\label{eq:2.8} 
\end{eqnarray} 
Then a set of fundamental bounds on ${\rm Var}[T]$, independent 
of the choice of the estimator $T(x)$, can be obtained 
by applying the operator $\sum_{i}\alpha^{i}
\partial_{i}$ to (\ref{eq:2.7}), letting $\alpha^{i}$ be 
arbitrary. By use of (\ref{eq:2.1}) and 
the Schwartz inequality for $L^{2}({\bf R}^{1})$, we obtain 
\begin{eqnarray}
g_{ij} {\rm Var}[T] \geq \frac{1}{4} 
\frac{\partial\tau}{\partial\theta^{i}}
\frac{\partial\tau}{\partial\theta^{j}} . 
\label{eq:2.9}
\end{eqnarray}
This matrix inequality is interpreted as saying that if 
we subtract the right side from the left, the result is 
nonnegative definite. It follows that if the 
random variables $\Theta^{i}$ $(i=1,\cdots,r)$ are unbiased 
estimators for the parameters $\theta^{i}$, satisfying 
\begin{eqnarray} 
\int_{-\infty}^{\infty} \rho(x,\theta) \Theta^{i}(x) {\rm d}x 
= \theta^{i} , 
\label{eq:2.10} 
\end{eqnarray} 
then the covariance matrix of the estimators is bounded by the 
inverse Fisher information matrix: 
\begin{eqnarray}
{\rm Cov}[\Theta^{i},\Theta^{j}] \geq \frac{1}{4}g^{ij} . 
\label{eq:2.11} 
\end{eqnarray} 

The Riemannian metric (\ref{eq:2.5}) introduced above can be 
used to define a distance measure between two distributions 
belonging to a given parametric 
family. This measure is invariant in the sense that it is 
unaffected by a reparameterisation of the distributions. The 
distance is calculated by integrating the infinitesimal line 
element ${\rm d}s$ along the geodesic connecting the two points in 
the statistical manifold ${\cal M}$, where 
\begin{eqnarray} 
{\rm d}s^{2} = \sum_{i,j} g_{ij} {\rm d}\theta^{i} {\rm d}
\theta^{j} . 
\label{eq:2.12} 
\end{eqnarray} 
The geodesics with respect to a given metric 
$g_{ij}$ are the solutions of the differential 
equation 
\begin{eqnarray} 
\frac{{\rm d}^{2}\theta^{i}}{{\rm d}u^{2}} + \Gamma^{i}_{jk} 
\frac{{\rm d}\theta^{j}}{{\rm d}u}
\frac{{\rm d}\theta^{k}}{{\rm d}u} = 0 
\label{eq:2.13} 
\end{eqnarray} 
for the curve $\theta^{i}(u)$ in ${\cal M}$, subject to 
the given boundary conditions at the two end points. Here, 
we have written 
\begin{eqnarray} 
\Gamma^{i}_{jk} = \frac{1}{2} g^{il} \left( \partial_{j}g_{kl} 
+ \partial_{k}g_{jl} - \partial_{l}g_{jk} \right) , 
\label{eq:2.14} 
\end{eqnarray} 
where $\partial_{i}=\partial/\partial\theta^{i}$, and the inverse 
metric $g^{ij}$, also appearing in (\ref{eq:2.11}), satisfies 
$g^{ij}g_{jk}=\delta^{i}_{k}$, where 
$\delta^{i}_{k}$ is the Kronecker delta. Note that in 
equations (\ref{eq:2.13}) and (\ref{eq:2.14}) above, and elsewhere 
henceforth in this article, we employ the standard Einstein 
summation convention on repeated indices. 

\begin{figure}[t] 
\centerline{ 
\psfig{file=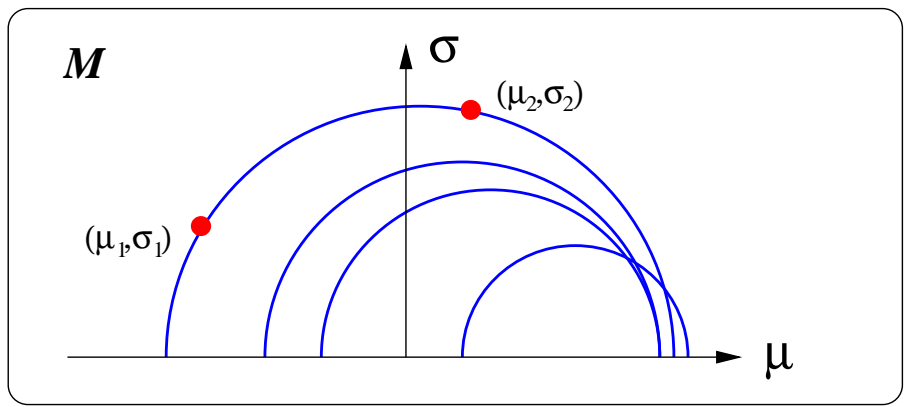,width=12cm,angle=0} 
}
\caption{{\it Geodesic curves for normal distributions}. The 
statistical manifold ${\cal M}$ in this case is the upper 
half plane parameterised by $\mu$ and $\sigma$. We have 
$-\infty<\mu<\infty$ and $0<\sigma<\infty$. The 
shortest path joining the two normal distributions 
${\cal N}(\mu_{1},\sigma_{1})$ and 
${\cal N}(\mu_{2},\sigma_{2})$ is given by the unique 
semi-circular arc through the given two points and centred 
on the boundary line $\sigma=0$. 
} 
\end{figure} 

Let us consider, as an explicit example, the manifold 
${\cal M}$ corresponding 
to the normal distributions ${\cal N}(\mu,\sigma)$ on 
${\bf R}^{1}$, with mean $\mu$ and standard deviation 
$\sigma$. For the parameterised density function we have 
\begin{eqnarray} 
\rho(x,\mu,\sigma) = \frac{1}{\sqrt{2\pi}\sigma}\exp\left( 
-\frac{(x-\mu)^{2}}{2\sigma^{2}} \right) . 
\label{eq:2.15} 
\end{eqnarray} 
A straightforward computation, making use of (\ref{eq:2.6}), 
gives 
\begin{eqnarray} 
{\rm d}s^{2} = \frac{1}{\sigma^{2}}({\rm d}\mu^{2} + 2 
{\rm d}\sigma^{2})  
\label{eq:2.16} 
\end{eqnarray} 
for the line element, which is defined on the upper half-plane 
$-\infty<\mu<\infty$, $0<\sigma<\infty$. The resulting Riemannian 
geometry is that of hyperbolic space, which is a homogeneous 
manifold with constant negative curvature. The geometry of this 
space has been studied extensively, and has many intriguing 
properties. For the distance function in the 
case of a pair of normal distributions 
${\cal N}(\mu_{1},\sigma_{1})$, 
${\cal N}(\mu_{2},\sigma_{2})$ we obtain 
\begin{eqnarray} 
D(\rho_{1},\rho_{2}) = \frac{1}{\sqrt{2}}\log
\frac{1+\delta_{1,2}}{1-\delta_{1,2}} , 
\label{eq:2.17} 
\end{eqnarray} 
where the function $\delta_{1,2}$, defined by 
\begin{eqnarray} 
\delta_{1,2} = \sqrt{\frac{(\mu_{2}-\mu_{1})^{2}
+2(\sigma_{2}-\sigma_{1})^{2}}{(\mu_{2}-\mu_{1})^{2}+2
(\sigma_{2}+\sigma_{1})^{2}}} , 
\label{eq:2.18} 
\end{eqnarray} 
lies between 0 and 1. The geodesics, in particular, are given 
in general by semi-circular arcs 
centred on the boundary line $\sigma=0$ (this line itself is 
not part of the manifold ${\cal M}$). An exceptional situation 
arises when $\mu_{1}=\mu_{2}$, for which the geodesic is a 
straight line given by constant $\mu$, and we have 
\begin{eqnarray} 
D(\rho_{1},\rho_{2}) = \frac{1}{\sqrt{2}}\left|\log
\frac{\sigma_{1}}{\sigma_{2}}\right| . 
\label{eq:2.19} 
\end{eqnarray} 
We refer the reader to Burbea (1986), where metric and 
distance computations have been carried out explicitly for 
other families of distributions. 

\section{Discount bond densities} 

Our goal now is to make use of the analysis presented in the 
previous section to construct a natural metric on the space of 
yield curves. In doing so we shall take advantage of a remarkable 
`probabilistic' characterisation of discount bonds, which we here 
proceed to describe. 

Let $t=0$ denote the present, and $P_{0T}$ a smooth 
family of discount bonds, where $T$ is the maturity date 
$(0\leq T<\infty)$. For positive interest we require 
\begin{eqnarray} 
0<P_{0T}\leq1, \ \ \ \frac{\partial}{\partial T} 
P_{0T} < 0 , 
\label{eq:3.1} 
\end{eqnarray} 
and we assume that $P_{0T}\rightarrow0$ as $T$ goes to 
infinity. A term structure that satisfies these conditions 
will be said to be `admissible'. These conditions can, 
in fact, be relaxed slightly: 
$P_{0T}$ need not be strictly smooth, nor strictly decreasing; 
but for most of the present discussion we shall 
stick with the assumptions indicated. 

\begin{figure}[t] 
\centerline{ 
\psfig{file=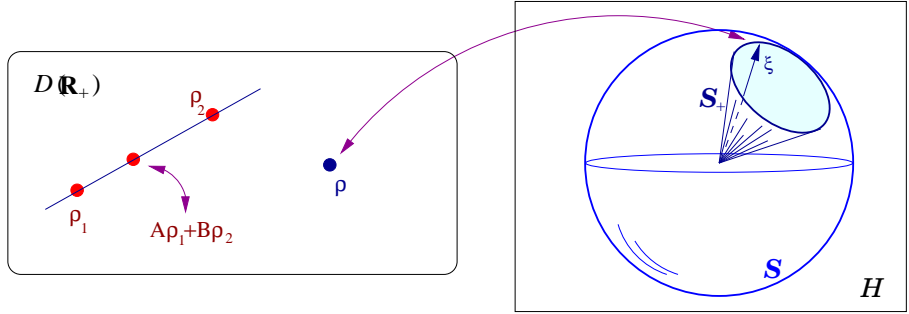,width=12cm,angle=0} 
}
\caption{{\it The system of admissible term structures}. 
A smooth positive interest term structure can be regarded 
as a point in ${\cal D}({\bf R}_{+}^{1})$, the convex 
space consisting of smooth density functions on 
${\bf R}_{+}^{1}$. The points of ${\cal D}({\bf R}_{+}^{1})$ 
are in one-to-one correspondence with rays lying in the 
positive orthant ${\cal S}_{+}$ of the unit sphere ${\cal S}$ 
in the Hilbert space ${\cal H}=L^{2}({\bf R}_{+}^{1})$. 
} 
\end{figure} 

The interesting point that arises here, of which we shall 
make extensive use in the discussions that follow, is that 
the discount function $P_{0T}$ can be viewed as a 
complementary probability distribution. In other words, we 
think of the maturity date as an abstract random variable $X$, 
and for its distribution we write 
\begin{eqnarray} 
{\bf P}[X<T] = 1-P_{0T} . 
\label{eq:3.2} 
\end{eqnarray} 
It should be clear that this can be done if and only if the 
positive interest rate conditions given in (\ref{eq:3.1}) hold. 
As a consequence we are able to embody the positive interest 
property in a fundamental way in the structure of the theory. 
Indeed, this basic economic property is essential if we wish 
to treat the yield curve consistently and naturally as a kind 
of mathematical object in its own right. Now let us introduce 
the function $\rho(T)$ defined by 
\begin{eqnarray} 
\rho(T) = -\frac{\partial}{\partial T} P_{0T} . 
\label{eq:3.3} 
\end{eqnarray} 
Clearly, we have $\rho(T)> 0$ and 
\begin{eqnarray} 
\int_{0}^{\infty} \rho(T) {\rm d}T = 1 , 
\label{eq:3.4} 
\end{eqnarray} 
from which we infer that $\rho(T)$ can be consistently viewed as 
a probability density function. It follows from the defining 
equation (\ref{eq:3.3}) that the term structure density 
$\rho(T)$ is the product of the instantaneous forward 
rate and the discount function itself. Now clearly if 
$\rho_{1}(T)$ and $\rho_{2}(T)$ are admissible term structure 
densities, and if $A$ and $B$ are nonnegative constants satisfying 
$A+B=1$, then $A\rho_{1}(T)+B\rho_{2}(T)$ is also an admissible 
term structure density. Putting these ingredients together, we 
see that the term structure of interest rates can be given the 
following general characterisation. 

\begin{prop} 
The system of admissible term structures is isomorphic to the 
convex space ${\cal D}({\bf R}_{+}^{1})$ of smooth density 
functions on the positive real line. 
\label{prop:1}
\end{prop} 

At first glance it may seem odd to think of the discount function 
in this manner. However, it gives us the advantage of being able to 
apply the tools of information geometry in an unexpected way, as 
we indicate in what follows. 

In particular, there is a one-to-one map from the space 
${\cal D}({\bf R}_{+}^{1})$ of such term structure densities to 
the positive orthant ${\cal S}_{+}$ of the unit sphere ${\cal S}$ 
in the Hilbert space ${\cal H}$, as indicated in Figure 2. 
Therefore, given two yield 
curves we can calculate the distance between them. This can be 
carried out either in a nonparametric sense, by use of the 
Bhattacharyya spherical distance, or in a parametric sense, by 
use of the Fisher-Rao distance. In the former case first we 
calculate the corresponding term structure densities $\rho_{1}(T)$ 
and $\rho_{2}(T)$. These are then mapped to ${\cal S}_{+}$ by 
taking the square-roots, and their distance $\phi(\rho_{1},
\rho_{2})$ is given by 
\begin{eqnarray} 
\phi(\rho_{1},\rho_{2}) = \cos^{-1} \int_{0}^{\infty} 
\sqrt{\rho_{1}(T)\rho_{2}(T)}{\rm d}T .
\label{eq:3.41} 
\end{eqnarray} 
In the parametric case we regard the given parametric family of 
yield curves as defining a statistical model ${\cal M}\subset 
{\cal S}_{+}$, and the distance between the two yield curves 
within the given family is then defined by the Fisher-Rao metric.

\section{Flat term structures} 

To provide some illustrations of the principles set 
forth in the previous section we consider here properties of 
yield curves for which the term structure is {\sl flat}. Such 
yield curves, which are of various types, are on the whole 
too simple for use in practical modelling. Nevertheless, 
they are of interest as examples, because many of the relevant 
computations can be carried out explicitly. 

In this connection we begin by introducing a representation of 
the discount function as a Laplace transform 
\begin{eqnarray} 
P_{0T} = \int_{0}^{\infty} e^{-rT} \psi(r) {\rm d}r 
\label{eq:4.1} 
\end{eqnarray}  
for some function $\psi(r)$. Thus we think of the discount 
function $P_{0T}$ as being given by a weighted superposition 
of elementary discount functions, each of the form $e^{-rT}$ 
for some value of $r$. Taking the limit $T\rightarrow0$, we 
find that $\psi(r)$ must satisfy $\int_{0}^{\infty} \psi(r) 
{\rm d}r = 1$. 
In general the inverse Laplace transform $\psi(r)$ need not 
be positive. However, if we restrict our consideration to 
nonnegative functions, then $\psi(r)$ can be interpreted as 
a density function, and by various 
choices of $\psi(r)$ we are led to some interesting 
candidates for term structures. 

First we consider the case where $\psi(r)$ is 
a Dirac $\delta$-function concentrated at a point, that is, 
$\psi(r)=\delta(r-R)$. A direct substitution gives 
$P_{0T}=\exp(-RT)$, 
corresponding to a `flat' term structure with a continuously 
compounded rate $R$ for each value of the maturity date $T$. 
If the density function $\psi(r)$ is given by an exponential 
distribution $\psi(r)=\tau\exp(-\tau r)$, with parameter 
$\tau$, then one sees that $\tau$ must have dimensions of 
time, and a short calculation gives $P_{0T}=\tau/(\tau+T)$, 
which also corresponds to a flat term structure, in this 
case with a simple percentage yield of $\tau^{-1}$ for all 
maturities. We see that the characteristic time-scale $\tau$ 
allows us to define an interest rate $R=\tau^{-1}$, which 
turns out to be the characteristic interest rate of the 
resulting structure, and we can write 
$P_{0T}=1/(1+RT)$ for the discount function. 

We note that flatness is not a completely unambiguous notion, 
because having a uniform continuously compounded yield for all 
maturities is not the same thing as having a uniform simple 
yield for all maturities. Both define plausible albeit quite 
distinct systems of discount bonds. This example illustrates 
how by superposing term structures of the elementary form 
$\exp(-RT)$ for various maturities, we can obtain other 
reasonable looking and well behaved term structures. 
We mention one more example, which contains the 
previous two examples as special cases. Consider the 
standard gamma distribution, with parameters $\kappa$ and 
$\lambda$, defined for nonnegative values of $r$ by the 
density function 
\begin{eqnarray} 
\psi(r) = \frac{1}{\Gamma(\kappa)} \lambda^{\kappa} 
r^{\kappa-1} \exp(-\lambda r) . 
\label{eq:4.6} 
\end{eqnarray} 
In this case, we can verify that the resulting system of 
discount bonds is given by 
\begin{eqnarray}  
P_{0T} = \left( \frac{\lambda}{\lambda+T} \right)^{\kappa} , 
\label{eq:4.7} 
\end{eqnarray} 
which assumes a more recognisable form if we set 
$\lambda=\kappa\tau$, where $\tau$ again defines a characteristic 
time scale, and $\kappa$ is a dimensionless number. Then we have 
\begin{eqnarray}  
P_{0T} = \left(1+\frac{RT}{\kappa} \right)^{-\kappa} , 
\label{eq:4.8} 
\end{eqnarray} 
where $R=\tau^{-1}$. The system of discount bonds 
arising here can also be interpreted as a flat term structure, 
in this case with a constant annualised rate of interest $R$ 
assuming compounding at the frequency $\kappa$ over the life 
of each bond ($\kappa$ need not be an integer). It is not 
difficult to check that for $\kappa=1$ this reduces to the 
case of a flat rate on the basis of a simple yield, whereas 
in the limit $\kappa\rightarrow\infty$ we recover the case of 
a flat rate on the basis of continuous compounding. 

Now we shall apply the ideas of statistical geometry to make 
comparisons between various term structures of the form 
(\ref{eq:4.8}). For density function $\rho(T)=-\partial_{T}
P_{0T}$ in this case we obtain 
\begin{eqnarray} 
\rho(T,R) = R\left( 1+\frac{RT}{\kappa}\right)^{-(\kappa+1)}.  
\label{eq:4.9} 
\end{eqnarray} 
Here we find it convenient to label the density function 
by the flat rate $R$. Note that in the limit 
$\kappa\rightarrow\infty$ we have 
$\rho(T,R)\rightarrow Re^{-RT}$. First consider 
the nonparametric separation between 
different term structures in this model via spherical distance 
of Bhattacharyya given in formula (\ref{eq:3.41}), where in the 
present example we write $\rho_{i}(T)=\rho(T,R_{i})$ for 
$i=1,2$. A direct integration leads to the expression 
\begin{eqnarray} 
\phi(\rho_{1},\rho_{2}) = \cos^{-1}\left( 
\frac{\sqrt{R_{1}R_{2}}}{R_{1}-R_{2}}\log\frac{R_{1}}{R_{2}}
\right) 
\label{eq:4.11} 
\end{eqnarray} 
for the distance when $\kappa=1$, whereas in the limit 
$\kappa\rightarrow\infty$ (continuous compounding) we have 
\begin{eqnarray} 
\phi(\rho_{1},\rho_{2}) = \cos^{-1}\left( 
\frac{2\sqrt{R_{1}R_{2}}}{R_{1}+R_{2}}\right) . 
\label{eq:4.12} 
\end{eqnarray} 
It is interesting to observe that the bracketed term in 
(\ref{eq:4.12}) is given by the ratio of the geometric and 
arithmetic means of the two rates. 

Alternatively, we can view (\ref{eq:4.9}) as a parametric 
family of distributions, parameterised by the flat rate $R$. 
Then it is natural to consider the Fisher-Rao distance 
between the two term structures characterised by $R_{1}$ and 
$R_{2}$. A straightforward calculation then leads to a 
simple distance formula given by 
\begin{eqnarray} 
D(R_{1},R_{2}) = \sqrt{\frac{\kappa}{\kappa+2}}\log
\frac{R_{2}}{R_{1}} , 
\label{eq:4.13} 
\end{eqnarray} 
where we have assumed $R_{2}\geq R_{1}$. 

\section{Interest rate dynamics} 

The formalism we have developed so far is essentially a static 
one, set in the present. Now we turn to the problem of developing 
a dynamical theory of interest rates. The idea is that, at each 
instant of time, the yield curve is characterised by a term 
structure density according to the scheme described in the previous 
sections. Then, as time passes, the density function evolves 
randomly. As a consequence we obtain a measure-valued process. 
In particular, we obtain a process on ${\cal D}({\bf R}_{+}^{1})$. 
Our goal in this section is to determine a set of conditions on 
this process necessary and sufficient to ensure that the resulting 
interest rate dynamics will be arbitrage-free. 

We shall assume the reader is familiar with the general theory 
of interest rate dynamics as laid out, for example, in Carverhill 
(1994), Rogers (1994), Hughston (1996), Baxter (1997), Musiela 
$\&$ Rutkowski (1997), Brody (2000) or Hunt $\&$ Kennedy (2000). 
For the general discount bond dynamics, let us write 
\begin{eqnarray} 
{\rm d}P_{tT} = \mu_{tT}{\rm d}t + \Sigma_{tT}\cdot {\rm d}
W_{t} , 
\label{eq:5.1} 
\end{eqnarray} 
where $\mu_{tT}$ and $\Sigma_{tT}$ are the {\sl absolute drift 
and absolute volatility processes}, respectively, for a bond 
with maturity $T$. Here, 
$W_{t}$ is a vector Brownian motion, and $\Sigma_{tT}$ is a 
vector process, and there is an inner product implied between 
$\Sigma_{tT}$ and ${\rm d}W_{t}$, signified by a dot. We need not 
specify the dimensionality of the Brownian motion, which might 
be infinite, and indeed in some respects the infinite dimensional 
setting is the most natural one. 
In fact, it suffices for our purposes merely to 
assume that $P_{tT}$ is a one-parameter family of 
continuous semi-martingales on the given probability space, with 
respect to the given filtration. However, for simplicity of 
exposition we shall stick to the case where the relevant 
stochastic basis is generated by a multidimensional Brownian 
motion. Here, as in Flesaker $\&$ Hughston (1997a,b), we 
regard the discount bond dynamics as the natural 
starting position, rather than, say, the instantaneous forward 
rate dynamics (Heath {\it et al.} 1992), which we need not 
consider here directly. We shall assume nevertheless, as in the 
HJM framework, that the 
processes $\mu_{tT}$ and $\Sigma_{tT}$ are both smooth in the 
variable $T$, and that sufficiently strong technical 
conditions are in place to ensure that the instantaneous 
forward rate processes are semimartingales. 

In order to extend the analysis of the previous section it is 
convenient to introduce what is sometimes conveniently referred 
to as the `Musiela parameterisation', given by 
\begin{eqnarray} 
B_{tx} = P_{t,t+x} , 
\label{eq:5.2} 
\end{eqnarray} 
where $T=t+x$ represents the maturity date of the bond, and 
hence $x$ is the time left until maturity. Thus $B_{tx}$ is the 
value at time $t$ of a discount bond that has $x$ years left to 
mature. This choice of parameterisation has already been shown 
to be useful in the geometric analysis of interest rates 
(Bj\"ork $\&$ Svensson 1999, Bj\"ork $\&$ Christensen 1999, 
Bj\"ork $\&$ Gombani 1999, Bj\"ork 2000). We note that 
$B_{t0}=1$ for all 
$t$, and that $B_{tx} \rightarrow 0$ as $x\rightarrow\infty$. It 
follows that 
\begin{eqnarray} 
\rho_{t}(x) = -\frac{\partial}{\partial x}B_{tx} 
\label{eq:5.4} 
\end{eqnarray} 
is a measure-valued process in the sense that, for each 
value of $t$ the random function $\rho_{t}(x)$ satisfies 
$\rho_{t}(x)>0$ and the normalisation condition 
\begin{eqnarray} 
\int_{0}^{\infty} \rho_{t}(x) {\rm d}x = 1 . 
\label{eq:5.5} 
\end{eqnarray} 
Here we have chosen the 
notation $\rho_{t}(x)$ that makes the $x$ dependence more 
prominent, to emphasise the fact that, for each value of $t$, 
and conditional on information given up to time $t$, 
$\rho_{t}(x)$ is a density function, though we might have 
written $\rho_{tx}$ instead. As a consequence $\rho_{t}(x)$ 
describes a process on ${\cal D}({\bf R}_{+}^{1})$. By 
consideration of (\ref{eq:5.1}) and (\ref{eq:5.2}) we deduce 
for the dynamics of $B_{tx}$ that  
\begin{eqnarray} 
{\rm d}B_{tx} = \left. {\rm d}P_{tT}\right|_{T=t+x} + 
\frac{\partial}{\partial x}B_{tx}{\rm d}t ,  
\label{eq:5.6} 
\end{eqnarray} 
and thus, by use of (\ref{eq:5.1}), that 
\begin{eqnarray} 
{\rm d}B_{tx} = \left(\mu_{t,t+x}+\partial_{x}B_{tx}\right) 
{\rm d}t + \Sigma_{t,t+x}\cdot {\rm d}W_{t} ,  
\label{eq:5.7} 
\end{eqnarray} 
where $\partial_{x}=\partial/\partial x$. Differentiating this 
expression with respect to $x$ and introducing the measure-valued 
process $\rho_{t}(x)$ according to formula (\ref{eq:5.4}) we 
therefore obtain 
\begin{eqnarray} 
{\rm d}\rho_{t}(x) = \left(-\partial_{x}\mu_{t,t+x}+\partial_{x}
\rho_{t}(x)\right){\rm d}t - \partial_{x} \Sigma_{t,t+x}\cdot 
{\rm d}W_{t} .  
\label{eq:5.8} 
\end{eqnarray} 
A further simplification is then achieved by introducing the 
notation 
\begin{eqnarray} 
\beta_{tx} = - \partial_{x}\mu_{t,t+x}
\label{eq:5.9} 
\end{eqnarray} 
and 
\begin{eqnarray} 
\omega_{tx} = -\partial_{x} \Sigma_{t,t+x} , 
\label{eq:5.10} 
\end{eqnarray} 
which gives us 
\begin{eqnarray} 
{\rm d}\rho_{t}(x) = \left( \beta_{tx} + \partial_{x}
\rho_{t}(x)\right){\rm d}t + \omega_{tx}\cdot {\rm d}W_{t} .  
\label{eq:5.11} 
\end{eqnarray} 

In the foregoing discussion we have not yet imposed the 
arbitrage-free condition. This is given by the drift constraint 
\begin{eqnarray} 
\mu_{tT} = r_{t} P_{tT} + \Sigma_{tT}\cdot \lambda_{t} , 
\label{eq:5.12} 
\end{eqnarray} 
where $\lambda_{t}$ is the process for the market price of risk. 
We note that $\lambda_{t}$, like $\Sigma_{tT}$, is a vector 
process. However, $\lambda_{t}$ does not depend on the maturity 
$T$. The absence of arbitrage ensures the existence of 
$\lambda_{t}$. For our purposes we do not need to insist that 
the bond market is complete: all we require is the existence 
of a pricing kernel, or equivalently the existence of a 
self-financing 
`natural numeraire' portfolio with value process $N_{t}$ such 
that $P_{tT}/N_{t}$ is a martingale for each value of $T$ 
(cf. Flesaker $\&$ Hughston 1997c). The numeraire process 
satisfies  
\begin{eqnarray} 
\frac{{\rm d}N_{t}}{N_{t}} = (r_{t}+\lambda_{t}^{2}){\rm d}t 
+ \lambda_{t}\cdot {\rm d}W_{t} , 
\label{eq:5.121} 
\end{eqnarray} 
and the corresponding pricing kernel is given by $1/N_{t}$. 
As a consequence of the constraint (\ref{eq:5.12}) we then 
have 
\begin{eqnarray} 
\mu_{t,t+x} = r_{t} B_{tx} + \Sigma_{t,t+x}\cdot \lambda_{t} , 
\label{eq:5.13} 
\end{eqnarray} 
and therefore, by differentiation of this expression with 
respect to $x$, we obtain 
\begin{eqnarray} 
\beta_{tx} = r_{t}\rho_{t}(x) + \omega_{tx}\cdot\lambda_{t} . 
\label{eq:5.14} 
\end{eqnarray} 
Inserting (\ref{eq:5.14}) in (\ref{eq:5.11}) we are thus able to 
express the dynamics of the density function $\rho_{t}(x)$ in the 
form 
\begin{eqnarray} 
{\rm d}\rho_{t}(x) = \left( r_{t}\rho_{t}(x) + 
\partial_{x}\rho_{t}(x) \right) {\rm d}t + \omega_{tx} 
\cdot \left( {\rm d}W_{t}+\lambda_{t}{\rm d}t\right) . 
\label{eq:5.15} 
\end{eqnarray} 

Before proceeding further, let us verify, as a consistency check, 
that the dynamics given by (\ref{eq:5.15}) preserves the 
normalisation condition on $\rho_{t}(x)$, given by (\ref{eq:5.5}). 
Integrating the right hand side of (\ref{eq:5.15}) with respect to 
$x$ and equating the drift and volatility terms separately to zero 
leads to the relations 
\begin{eqnarray} 
r_{t} + \int_{0}^{\infty} \partial_{x}\rho_{t}(x){\rm d}x = 0  
\label{eq:5.16} 
\end{eqnarray} 
and 
\begin{eqnarray} 
\int_{0}^{\infty} \omega_{tx} {\rm d}x = 0 , 
\label{eq:5.17} 
\end{eqnarray} 
which must hold for all $t$. Condition (\ref{eq:5.16}) is satisfied 
because $\rho_{t}(x)\rightarrow0$ as $x\rightarrow\infty$ and 
\begin{eqnarray} 
\rho_{t}(0) = r_{t} . 
\label{eq:5.18} 
\end{eqnarray} 
Condition (\ref{eq:5.17}) is satisfied because, by definition, 
we have 
$\omega_{tx}=-\partial_{x}\Sigma_{t,t+x}$, and the absolute 
volatility $\Sigma_{t,t+x}$ vanishes both as $x\rightarrow0$ (a 
maturing bond has a definite value and thus has no absolute 
volatility), and as $x\rightarrow\infty$ (a bond with infinite 
maturity has no value, and hence no absolute volatility).  

Summing up matters so far, we see that in (\ref{eq:5.15}) we are 
able to cut the standard HJM arbitrage-free interest rate dynamics 
in the form of a measure-valued process $\rho_{t}(x)$ subject to 
the constraints (\ref{eq:5.16}) and (\ref{eq:5.17}). At first 
glance the role of the short rate $r_{t}$ in (\ref{eq:5.15}) seems 
anomalous, because it might appear that this has to be specified 
separately. However, by virtue of ({\ref{eq:5.18}) we can 
incorporate $r_{t}$ directly into the dynamics of $\rho_{t}(x)$. 

In fact, there is another way of expressing (\ref{eq:5.15}) which 
is very suggestive, and ties in naturally with the Hilbert space 
approach to dynamics introduced in \S 7. First 
we note that (\ref{eq:5.16}) can be rewritten in the form 
\begin{eqnarray} 
r_{t} = -\int_{0}^{\infty} \rho_{t}(x) \partial_{x}\ln \rho_{t}(x) 
{\rm d}x .
\label{eq:5.19} 
\end{eqnarray} 
In other words, $r_{t}$ is minus the expectation of the gradient 
of the log-likelihood function. Here the expectation is taken 
with respect to $\rho_{t}(x)$ itself. Writing $E_{\rho}$ for this 
abstract expectation, we have 
\begin{eqnarray} 
{\rm d}\rho_{t}(x) = \rho_{t}(x) \left( \partial_{x}\ln\rho_{t}(x) 
- E_{\rho}[\partial_{x}\ln\rho_{t}(x)]\right) 
{\rm d}t + \omega_{tx}\cdot {\rm d}W_{t}^{*} , 
\label{eq:5.20} 
\end{eqnarray} 
where ${\rm d}W_{t}^{*}={\rm d}W_{t}+\lambda_{t}{\rm d}t$. 
We note that $W_{t}^{*}$ has the interpretation of being a 
Brownian motion with respect to the risk-neutral measure 
associated with the given pricing kernel. In the risk-neutral 
measure, for which the term 
involving $\lambda_{t}$ effectively disappears, the remaining 
drift for $\rho_{t}(x)$ is determined by the deviation of 
$\partial_{x}\ln\rho_{t}(x)$ from its abstract mean. 

Let us now examine more closely the volatility term 
$\omega_{tx}$ appearing in (\ref{eq:5.20}), with a view to gaining 
a better understanding of the significance of the volatility 
constraint 
(\ref{eq:5.17}). Because $\rho_{t}(x)$ must remain positive 
for all values of $x$, the coefficient of 
${\rm d}W_{t}^{*}$ in (\ref{eq:5.20}) must be of the form 
\begin{eqnarray} 
\omega_{tx} = \rho_{t}(x) \sigma_{tx} 
\label{eq:5.22} 
\end{eqnarray}
for some bounded process $\sigma_{tx}$, to ensure that 
$\omega_{tx}$ dies off appropriately for values of $x$ such 
that $\rho_{t}(x)$ approaches zero. As a consequence, we can 
write (\ref{eq:5.15}) in the quasi-lognormal form 
\begin{eqnarray} 
\frac{{\rm d}\rho_{t}(x)}{\rho_{t}(x)} = (r_{t}+\partial_{x}
\ln\rho_{t}(x)){\rm d}t + \sigma_{tx}\cdot 
{\rm d}W_{t}^{*}, 
\label{eq:5.23} 
\end{eqnarray} 
and for the constraint (\ref{eq:5.17}) we have 
\begin{eqnarray} 
E_{\rho}[\sigma_{tx}] = 0, 
\label{eq:5.24}
\end{eqnarray} 
which can be satisfied by writing 
\begin{eqnarray} 
\sigma_{tx} = \nu_{tx} - E_{\rho}[\nu_{tx}] , 
\label{eq:5.25}
\end{eqnarray} 
where $\nu_{tx}$ is an exogenously specifiable unconstrained 
process. Here, for any process $A_{tx}$ we define 
$E_{\rho}[A_{tx}] = \int_{0}^{\infty}\rho_{t}(x)A_{tx}{\rm d}x$. 
The results established above can then be summarised as follows. 

\begin{prop} 
The general admissible term structure evolution based on the 
information set generated by a multidimensional Brownian motion 
$W_{t}$ is given by a measure-valued process $\rho_{t}(x)$ in 
${\cal D}({\bf R}_{+}^{1})$ satisfying 
\begin{eqnarray} 
\frac{{\rm d}\rho_{t}(x)}{\rho_{t}(x)} &=& \left( 
\partial_{x}\ln\rho_{t}(x)-E_{\rho}[\partial_{x}\ln\rho_{t}(x)]
\right) {\rm d}t \nonumber \\ 
& & + \left( \nu_{tx}-E_{\rho}[\nu_{tx}]\right) 
\cdot \left( {\rm d}W_{t}+\lambda_{t}{\rm d}t \right) , 
\label{eq:5.26} 
\end{eqnarray} 
where the processes $\lambda_{t}$ and $\nu_{tx}$ are 
specified exogenously, along with the initial term structure 
density $\rho_{0}(x)$. 
\end{prop} 

An advantage of the particular expression (\ref{eq:5.26}) 
given for the dynamics above is that the preservation of the 
normalisation condition on $\rho_{t}(x)$ is evident by 
inspection, because this is equivalent to the relation 
\begin{eqnarray} 
E_{\rho}\left[ \frac{{\rm d}\rho_{t}(x)}{\rho_{t}(x)} 
\right] = 0 . 
\label{eq:5.27} 
\end{eqnarray} 
An alternative expression for (\ref{eq:5.26}), which brings 
out more explicitly the nonlinearities in the dynamics, is 
given by 
\begin{eqnarray} 
{\rm d}\rho_{t}(x) &=& \left(\partial_{x}\rho_{t}(x)+
\rho_{t}(0)\rho_{t}(x)\right){\rm d}t \nonumber \\ 
& & + \rho_{t}(x)\left( 
\nu_{tx}-\int_{0}^{\infty}\rho_{t}(y)\nu_{ty}{\rm d}y 
\right) \cdot {\rm d}W^{*}_{t}, 
\label{eq:5.28} 
\end{eqnarray} 
where ${\rm d}W^{*}_{t}={\rm d}W_{t}+\lambda_{t}{\rm d}t$ as 
defined earlier. 

\section{Principal moment analysis} 

The characterisation of the yield curve as an abstract probability 
density enables us to develop a rigourous analogue of the classical 
`principal component' analysis often used in the study of yield 
curve dynamics. To this end we let $\rho_{t}(x)=-\partial_{x}
P_{t,t+x}$ be the density process associated with an admissible 
family of discount bond prices, and define the moment processes 
\begin{eqnarray}
{\bar x}_{t} = \int_{0}^{\infty} x \rho_{t}(x) {\rm d}x 
\label{eq:66.1} 
\end{eqnarray}
and 
\begin{eqnarray}
{\bar x}_{t}^{\prime(n)} = \int_{0}^{\infty} x^{n} \rho_{t}(x) {\rm d}x 
\label{eq:66.2} 
\end{eqnarray}
for $n\geq2$, along with the central moment processes 
\begin{eqnarray}
{\bar x}_{t}^{(n)} = \int_{0}^{\infty} (x-{\bar x}_{t})^{n}  
\rho_{t}(x) {\rm d}x . 
\label{eq:66.3} 
\end{eqnarray}

It is important to note that in some cases the relevant moments 
may not exist. For example, in the case of a continuously 
compounded flat yield curve given at $t=0$ by the density 
function $\rho_{0}(x)=R{\rm e}^{-Rx}$, we have ${\bar x}_{0}=R^{-1}$, 
${\bar x}_{0}^{(2)} = R^{-2}$, ${\bar x}_{0}^{(3)}=3R^{-3}$, and 
${\bar x}_{0}^{(4)}=9R^{-4}$ for the first four central moments. 
On the other hand, in the example of the simple flat term 
structure for which $\rho_{0}(x)=R/(1+Rx)^{2}$ we find that none 
of the moments exist, on account of the fatness of the tail of 
the distribution. In fact, for the flat rate term structures with 
compounding frequency $\kappa$ the moments exist only up to order 
$\kappa-1$. 

The first four moments, if they exist, are the mean, variance, 
skewness and kurtosis of the distribution of the abstract 
random variable $X$ characterising the yield curve, and we refer 
to these (and other) moments as the 
`principal moments' of the given term structure. 
At $t=0$ the mean ${\bar x}_{0}$ determines a characteristic 
time-scale associated with the given term structure, and 
its inverse $1/{\bar x}_{0}$ can be thought of as an associated 
characteristic yield. The difference ${\bar x}_{0}^{(2)}
-({\bar x}_{0})^{2}$ then measures the departure of the given 
term structure from flatness on a continuously compounded basis. 
This is on account of the fact that in the case of an exponential 
distribution the variance is given by the square of the mean. 

It is legitimate to conjecture that for some purposes the 
specification of, e.g., the first three or four moments will be 
sufficient to provide an accurate representation of the term 
structure. One way of implementing this idea is to introduce the 
entropy $S_{\rho}$ of the given distribution, defined by 
\begin{eqnarray}
S_{\rho} = -\int_{0}^{\infty}\rho(x)\ln\rho(x){\rm d}x . 
\label{eq:66.4}
\end{eqnarray}
Because $\rho(x)$ has dimensions of inverse time, $S_{\rho}$ is 
defined only up to an overall additive constant. Therefore, the 
difference of the entropies associated with two yield curves has 
an invariant significance. 

For yield curve calibration we 
propose that $\rho(x)$ should be chosen such that $S_{\rho}$ is 
maximised subject to the constraints of the data available. For 
example, if we are given as data only the mean ${\bar x}_{0}$, 
then the maximum entropy term structure is $\rho_{0}(x)=Re^{-Rx}$, 
where $R=1/{\bar x}_{0}$. 

It is also of great interest to study the dynamics of the 
principal characteristics in the case of a general admissible 
arbitrage-free term structure. We examine here, in particular, 
the mean and the variance processes. For this purpose we 
introduce a simplified notation $v_{t}={\bar x}_{t}^{(2)}$ for 
the variance process, i.e., 
\begin{eqnarray}
v_{t}=\int_{0}^{\infty}x^{2}\rho_{t}(x){\rm d}x - 
({\bar x}_{t})^{2} , 
\label{eq:66.5}
\end{eqnarray}
where the mean process ${\bar x}_{t}$ is given as in (\ref{eq:66.1}). 
We assume that both $\rho_{t}(x)$ and the discount bond volatility 
$\Sigma_{t,t+x}$ fall off to zero sufficiently rapidly to ensure that 
$\lim_{x\rightarrow\infty}x^{n}\rho_{t}(x)=0$ and 
$\lim_{x\rightarrow\infty}x^{n}\Sigma_{t,t+x}=0$ for $n=1,2$, and 
that the integrals $\int_{0}^{\infty}x^{n}\rho_{t}(x){\rm d}x$ and 
$\int_{0}^{\infty}x^{n-1}\Sigma_{t,t+x}{\rm d}x$ exist for 
$n=1,2$. 
A straightforward calculation then leads us to the following 
conclusion: 

\begin{prop} 
The first principal moment ${\bar x}_{t}$ of an admissible, 
arbitrage-free term structure satisfies the dynamical law 
\begin{eqnarray}
{\rm d}{\bar x}_{t} = (r_{t}{\bar x}_{t}-1){\rm d}t + 
{\bar\Sigma}_{t}\cdot{\rm d}W^{*}_{t} , 
\label{eq:66.6}
\end{eqnarray}
where ${\bar\Sigma}_{t}=\int_{0}^{\infty}\Sigma_{t,t+x}{\rm d}x$. 
\end{prop} 

{\it Proof}. Starting with (\ref{eq:5.23}) and (\ref{eq:66.1}) we 
have 
\begin{eqnarray}
{\rm d}{\bar x}_{t} &=& \int_{0}^{\infty}x{\rm d}\rho_{t}(x) 
{\rm d}x \nonumber \\ 
&=& \left( r_{t}{\bar x}_{t}+\int_{0}^{\infty}x\partial_{x}
\rho_{t}(x){\rm d}x\right) {\rm d}t - \left( \int_{0}^{\infty} 
x\partial_{x}\Sigma_{t,t+x}{\rm d}x\right)\cdot {\rm d}W_{t}^{*} 
\label{eq:66.61} 
\end{eqnarray}
by use of (\ref{eq:5.10}). Then, 
integrating by parts and using the assumed asymptotic behaviours 
for $\rho_{t}(x)$ and $\Sigma_{t,t+x}$, we obtain the desired 
result. \hspace*{\fill} $\diamondsuit$ 

We note that there is a critical level ${\bar x}_{t}^{*}$ for the 
first principal moment given by 
\begin{eqnarray}
{\bar x}_{t}^{*} = \frac{1}{r_{t}}(1 - 
\lambda_{t}\cdot{\bar\Sigma}_{t}).  
\label{eq:66.7}
\end{eqnarray}
When ${\bar x}_{t}>{\bar x}_{t}^{*}$ the drift of ${\bar x}_{t}$ 
is positive, and the drift increases further as ${\bar x}$ 
increases. On the other hand, when ${\bar x}_{t}<{\bar x}_{t}^{*}$, 
the drift of ${\bar x}_{t}$ is negative, and the drift decreases 
further as ${\bar x}_{t}$ decreases. For the variance process, we 
have: 

\begin{prop} 
The second principal moment $v_{t}$ of an admissible, 
arbitrage-free term structure satisfies the dynamical law 
\begin{eqnarray}
{\rm d}v_{t}=\left( r_{t}(v_{t}-{\bar x}_{t}^{2}) - 
{\bar\Sigma}_{t}^{2}\right) {\rm d}t + 2 \left( 
{\bar\Sigma}_{t}^{(1)}-{\bar x}_{t}{\bar\Sigma}_{t}\right)
\cdot {\rm d}W_{t}^{*} , 
\label{eq:66.8}
\end{eqnarray}
where ${\bar\Sigma}_{t}^{(1)}=\int_{0}^{\infty}x
\Sigma_{t,t+x}{\rm d}x$. 
\end{prop} 

{\it Proof}. Starting with formula (\ref{eq:66.5}) for $v_{t}$ 
we have 
\begin{eqnarray}
{\rm d}v_{t}=\int_{0}^{\infty} x^{2}{\rm d}\rho_{t}(x) {\rm d}x 
- {\rm d}({\bar x}_{t}^{2}) . 
\label{eq:66.81}
\end{eqnarray} 
For the first term we obtain 
\begin{eqnarray}
\int_{0}^{\infty}x^{2}\rho_{t}(x){\rm d}x &=& \left( r_{t} 
\int_{0}^{\infty} x^{2}\rho_{t}(x){\rm d}x+\int_{0}^{\infty} 
x^{2}\partial_{x}\rho_{t}(x){\rm d}x\right){\rm d}t \nonumber \\ 
& & - \left(\int_{0}^{\infty}x^{2}\partial_{x}\Sigma_{t,t+x} 
\right)\cdot {\rm d}W_{t}^{*} , 
\label{eq:66.82} 
\end{eqnarray}
where we have used (\ref{eq:5.10}) and (\ref{eq:5.23}). 
As a consequence of the assumed asymptotic behaviour of 
$\rho_{t}(x)$ and $\Sigma_{t,t+x}$, this becomes 
\begin{eqnarray}
\int_{0}^{\infty}x^{2}\rho_{t}(x){\rm d}x = \left( r_{t} 
{\bar x}_{t}^{\prime(2)} - 2{\bar x}_{t} \right){\rm d}t 
+ 2 {\bar \Sigma}_{t}^{(1)}\cdot {\rm d}W_{t}^{*} , 
\label{eq:66.83} 
\end{eqnarray}
after an integration by parts. For the second term in 
(\ref{eq:66.81}) we have 
\begin{eqnarray} 
{\rm d}({\bar x}_{t}^{2}) = 2{\bar x}_{t}{\rm d}{\bar x}_{t} + 
({\rm d}{\bar x}_{t})^{2} 
\label{eq:66.84} 
\end{eqnarray} 
by Ito's lemma, and thus 
\begin{eqnarray} 
{\rm d}({\bar x}_{t}^{2}) = \left( 2r_{t}{\bar x}_{t}^{2}-
2{\bar x}_{t}+\Sigma_{t}^{2}\right){\rm d}t + 2{\bar x}_{t} 
{\bar\Sigma}_{t}\cdot{\rm d}W_{t}^{*} 
\label{eq:66.85} 
\end{eqnarray} 
by use of Proposition 3. Combining (\ref{eq:66.83}) and 
(\ref{eq:66.85}), and using the definition (\ref{eq:66.5}) we 
obtain (\ref{eq:66.8}). \hspace*{\fill} $\diamondsuit$ 

In this case we recall that the difference $v_{t}-{\bar x}_{t}^{2}$ 
acts as a simple measure of the extent to which the distribution 
deviates from the `flat' term structure. As a consequence we see 
that the effect of the dynamics here is that the second principal 
moment of the term structure tends to increase, i.e., has a 
positive drift, providing 
$v_{t}-{\bar x}_{t}^{2}$ is already above the level given by  
\begin{eqnarray}
v_{t}-{\bar x}_{t}^{2} = \frac{1}{r_{t}}\left( 
{\bar\Sigma}_{t}^{2}-2\lambda_{t}\cdot ({\bar\Sigma}_{t}^{(1)}-
{\bar x}_{t}{\bar\Sigma}_{t})\right) . 
\label{eq:66.9}
\end{eqnarray}

\section{Hilbert space dynamics for term structures} 
 
Now that we have examined some of the advantages of expressing 
the arbitrage-free interest rate term structure dynamics as a 
randomly evolving density function, let us consider how we 
transform to the Hilbert space representation for density 
functions considered in \S 2. Denote by $\xi_{tx}$ the process 
for the square-root likelihood function, defined by  
\begin{eqnarray} 
\rho_{t}(x) = \xi_{tx}^{2} . 
\label{eq:6.1} 
\end{eqnarray} 
It follows then, by Ito's lemma, that 
\begin{eqnarray} 
{\rm d}\rho_{t}(x) = 2\xi_{tx}{\rm d}\xi_{tx} + 
({\rm d}\xi_{tx})^{2} , 
\label{eq:6.2} 
\end{eqnarray} 
and hence 
$({\rm d}\rho_{t}(x))^{2} = 4\xi_{tx}^{2}({\rm d}\xi_{tx})^{2}$. 
By rearranging (\ref{eq:6.2}) we thus obtain 
\begin{eqnarray} 
{\rm d}\xi_{tx} = \frac{1}{2\xi_{tx}} {\rm d}\rho_{t}(x) - 
\frac{1}{8\xi_{tx}^{3}}({\rm d}\rho_{t}(x))^{2} 
\label{eq:6.3} 
\end{eqnarray} 
for the dynamics of the process $\xi_{tx}$, and hence 
\begin{eqnarray} 
{\rm d}\xi_{tx} = \left( \partial_{x}\xi_{tx} + \frac{1}{2}
r_{t}\xi_{tx} - \frac{1}{8\xi_{tx}^{3}}\omega_{tx}^{2} 
\right) {\rm d}t + \frac{1}{2\xi_{tx}}\omega_{tx}\cdot 
{\rm d}W_{t}^{*} , 
\label{eq:6.4} 
\end{eqnarray} 
where $\omega^{2}_{tx}=\omega_{tx}\cdot\omega_{tx}$. Now 
suppose we define $\sigma_{tx}$ by the quotient 
\begin{eqnarray}
\sigma_{tx} = \frac{\omega_{tx}}{\xi_{tx}^{2}} , 
\label{eq:6.5} 
\end{eqnarray} 
as before, and set 
$\sigma_{tx}^{2}=\sigma_{tx}\cdot\sigma_{tx}$. Then the 
process for the square-root density $\xi_{tx}$ can be 
written in the form 
\begin{eqnarray} 
{\rm d}\xi_{tx} = \left( \partial_{x}\xi_{tx} + \half 
r_{t}\xi_{tx} - \octa \xi_{tx}\sigma_{tx}^{2} \right) 
{\rm d}t + \half \xi_{tx} \sigma_{tx}\cdot {\rm d}W_{t}^{*} . 
\label{eq:6.6} 
\end{eqnarray} 
We recall that the volatility process $\sigma_{tx}$ arising 
again in this connection, which is given more explicitly by 
the ratio 
\begin{eqnarray}
\sigma_{tx} = \frac{\partial_{x}\Sigma_{t,t+x}}
{\partial_{x}B_{tx}} , 
\label{eq:6.7} 
\end{eqnarray} 
can be specified exogenously, subject only to the 
condition that it has mean zero in the measure $\rho_{t}(x)$, 
which implies that $\sigma_{tx}$ can be written in the form 
(\ref{eq:5.25}). 

\begin{figure}[t] 
\centerline{ 
\psfig{file=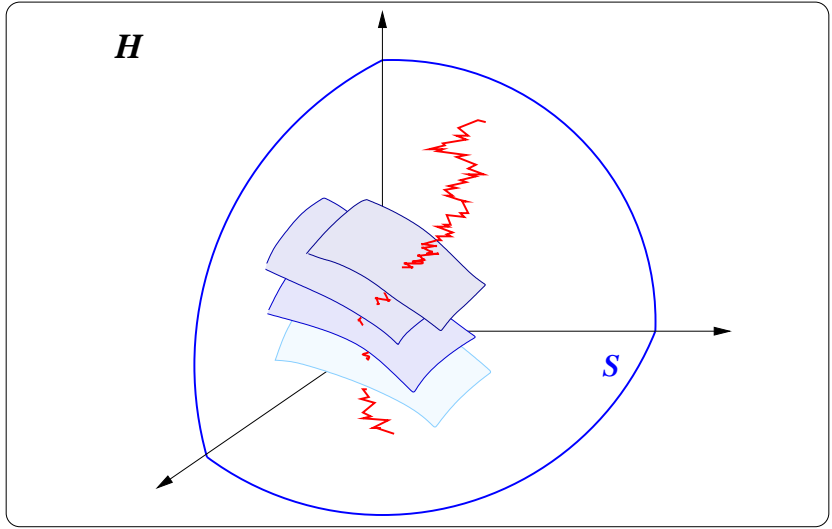,width=11cm,angle=0} 
}
 \caption{{\it Interest rate dynamics}. At each instant of 
time the interest rate term structure can be represented as 
a point on the positive orthant of the unit sphere 
${\cal S}$ in the Hilbert space ${\cal H}= 
L^{2}({\bf R}_{+}^{1})$. The associated arbitrage-free interest 
rate dynamics gives rise to a stochastic trajectory on this 
space, which is foliated by hypersurfaces corresponding to 
level values of the short-term interest rate. 
} 
\end{figure} 

We would now like to interpret the Hilbert space dynamics in 
equation (\ref{eq:6.6}) more directly in a geometrical fashion. 
For this purpose we find 
it expedient to introduce an index notation, using Greek 
letters to signify Hilbert space operations 
(cf. Brody $\&$ Hughston 1998). 

Thus if the function $\psi(x)$ is an element of 
${\cal H}=L^{2}({\bf R}_{+}^{1})$, we denote it by 
$\psi^{\alpha}$, and if $\varphi(x)$ belongs to the 
dual Hilbert space ${\cal H}^{*}$ we denote this by 
$\varphi_{\alpha}$. Furthermore, their inner product 
is written 
\begin{eqnarray}
\psi^{\alpha}\varphi_{\alpha} = \int_{0}^{\infty} 
\psi(x)\varphi(x) {\rm d}x . 
\label{eq:6.9} 
\end{eqnarray} 

There is a preferred symmetric quadratic form $g_{\alpha\beta}$ 
on ${\cal H}$, given by $g_{\alpha\beta}\psi^{\alpha}\psi^{\beta}
=\int_{0}^{\infty}(\psi(x))^{2}{\rm d}x$, which thus establishes 
an isomorphism between ${\cal H}$ and ${\cal H}^{*}$, given by 
$\psi^{\alpha}\rightarrow\psi_{\alpha}=g_{\alpha\beta}\psi^{\beta}$. 
Intuitively, one can think of $g_{\alpha\beta}$ as corresponding 
to the delta function $\delta(x,y)$, and then we have 
\begin{eqnarray}
g_{\alpha\beta} \psi^{\alpha}\varphi^{\beta} = \int_{0}^{\infty} 
\psi(x)\delta(x,y)\varphi(y) {\rm d}x {\rm d}y. 
\label{eq:6.10} 
\end{eqnarray} 
There are a number of Hilbert space technicalities 
that have to be considered for a complete exposition of the 
matter, but that is not our immediate concern. 

If $\xi(x)>0$ belongs to the positive orthant of 
$L^{2}({\bf R}^{1}_{+})$ then the corresponding indexed quantity 
$\xi^{\alpha}$ has the interpretation of a `state vector'. 
In that case we can think of symmetric quadratic forms as 
representing certain classes of random variables. 
The expectation of the random variable $H_{\alpha\beta}$ in the 
state $\xi^{\alpha}$ is 
\begin{eqnarray}
E_{\xi}[H] = \frac{H_{\alpha\beta}\xi^{\alpha}\xi^{\beta}}
{\xi_{\gamma}\xi^{\gamma}} . 
\label{eq:6.11} 
\end{eqnarray} 
Therefore, a state vector determines a mapping from random 
variables to real numbers, through (\ref{eq:6.11}). For a 
normalised state vector we have $\xi_{\alpha}\xi^{\alpha}=1$, 
although for some purposes it is convenient to relax the 
normalisation condition. In particular, we notice that the 
expectation (\ref{eq:6.11}) only depends on the direction of 
$\xi^{\alpha}$. 

Now suppose $\xi(x)$ is a positive function. In that case, the 
derivative $\partial_{x}$ can be thought of as a linear operator 
$D^{\alpha}_{\ \beta}$ on ${\cal H}$, and we have an 
endomorphism given by 
$\xi^{\alpha} \rightarrow D^{\alpha}_{\ \beta}\xi^{\beta}$. 
By making use of this, we can now interpret, in the language of 
Hilbert space geometry, the first two terms appearing in the 
drift in the dynamical equation (\ref{eq:6.6}). 

Let us begin 
by noting first that (\ref{eq:5.16}) can be rewritten in the 
form 
\begin{eqnarray}
\int_{0}^{\infty} \xi_{tx} \partial_{x} \xi_{tx} {\rm d}x 
= -\frac{1}{2} r_{t} . 
\label{eq:6.12} 
\end{eqnarray} 
This allows us to interpret the short term interest rate process 
$r_{t}$ in terms of the mean of the symmetric part of the operator 
$D^{\alpha}_{\ \beta}$ in the state $\xi_{t}^{\alpha}$, i.e., 
\begin{eqnarray} 
\frac{D_{\alpha\beta} \xi^{\alpha}_{t}\xi^{\beta}_{t}}
{g_{\alpha\beta}\xi^{\alpha}_{t}\xi^{\beta}_{t}} = 
-\frac{1}{2}r_{t} , 
\label{eq:6.13} 
\end{eqnarray} 
where $D_{\alpha\beta} = g_{\alpha\gamma}D^{\gamma}_{\ \beta}$. 
Therefore, if we let $D_{(\alpha\beta)}$ denote the symmetric 
part of the operator $D^{\alpha}_{\ \beta}$, then the abstract 
random variable in ${\cal H}$ corresponding to the short rate 
$r_{t}$ is given by $r_{\alpha\beta}=-2 D_{(\alpha\beta)}$. 
Similarly we can represent the abstract random variable $x$ 
for the time left until maturity in ${\cal H}$ by a symmetric 
matrix $X_{\alpha\beta}$. It is interesting to note that the 
random variables $X_{\alpha\beta}$ for the maturity date and 
$r_{\alpha\beta}$ for the short term interest rate are not 
`compatible'. Two random variables $A$ and $B$ are said to be 
compatible if the expression $\{\{A,C\},B\} - 
\{A,\{C,B\}\}$ vanishes for any random variable $C$, where 
$\{A,B\}=AB+BA$ denotes the anticommutator (Segal 1947). 
The lack of compatibility here indicates that the abstract 
probability system containing both $r_{\alpha\beta}$ and 
$X_{\alpha\beta}$ as random variables is not Kolmogorovian. 
However, the algebra of random variables generated by 
$X_{\alpha\beta}$ is Kolmogorovian. 

Now, let 
$\eta(x)$ be an arbitrary element of $L^{2}({\bf R}_{+}^{1})$, 
and let $\eta^{\alpha}$ be the corresponding Hilbert space 
vector. Then clearly we have 
\begin{eqnarray} 
\int_{0}^{\infty} \eta(x)\left[ \partial_{x} \xi_{tx} 
+ \half r_{t}\xi_{tx} \right] {\rm d}x = \eta_{\alpha} 
\left[ D^{\alpha}_{\ \beta}\xi^{\beta}_{t} - 
\left( 
\frac{D_{\beta\gamma}\xi^{\beta}_{t}\xi_{t}^{\gamma}}
{g_{\delta\epsilon}\xi_{t}^{\delta}\xi_{t}^{\epsilon}}
\right)
\xi^{\alpha}_{t} \right] . 
\label{eq:6.14} 
\end{eqnarray} 
In other words, the first two terms of the drift in (\ref{eq:6.6}) 
can be replaced by the expression ${\tilde D}^{\alpha}_{\ \beta}
\xi^{\beta}_{t}$, where 
\begin{eqnarray} 
{\tilde D}^{\alpha}_{\ \beta} = D^{\alpha}_{\ \beta} - 
\left( 
\frac{D_{\gamma\delta}\xi^{\gamma}_{t}\xi_{t}^{\delta}}
{g_{\gamma\delta}\xi_{t}^{\gamma}\xi_{t}^{\delta}} \right) 
\delta^{\alpha}_{\ \beta} ,  
\label{eq:6.15} 
\end{eqnarray} 
where $\delta^{\alpha}_{\ \beta}$ is the Kronecker delta. Clearly, 
we have ${\tilde D}_{\alpha\beta}\xi^{\alpha}\xi^{\beta}=0$. 

With this in mind, let us now proceed to the interpretation of the 
volatility process $\sigma_{tx}$. Again, $\sigma_{tx}$ has the 
character of a linear operator acting on $\xi_{tx}$, subject to 
the constraint $E_{\rho}[\sigma_{tx}]=0$. This can be consistently 
enforced if there exists a symmetric process 
$\nu_{t\alpha\beta}$ such that 
\begin{eqnarray} 
\int_{0}^{\infty} \eta(x) \xi_{tx} \sigma_{tx} 
{\rm d}x = \eta^{\alpha} 
\left( \nu_{t\alpha\beta}\xi^{\beta}_{t} - 
E_{\xi}[\nu_{t}] \xi_{t\alpha} \right) . 
\label{eq:6.16} 
\end{eqnarray} 
The symmetric operator-valued random process 
$\nu_{t\alpha\beta}$, whose existence is 
thus implied, is `primitive' in the sense that it is 
unconstrained and can be specified exogenously. If we write 
\begin{eqnarray} 
\sigma_{t\alpha\beta} = \nu_{t\alpha\beta} - 
E_{\xi}[\nu_{t}] g_{\alpha\beta} ,  
\label{eq:6.17} 
\end{eqnarray} 
we obtain 
\begin{eqnarray} 
\int_{0}^{\infty} \eta(x) \xi_{tx} \sigma_{tx} 
{\rm d}x = \eta_{\alpha} \sigma_{t\beta}^{\alpha} 
\xi^{\beta}_{t} , 
\label{eq:6.18} 
\end{eqnarray} 
and also 
\begin{eqnarray} 
\int_{0}^{\infty} \eta(x) \xi_{tx} \sigma_{tx}^2 
{\rm d}x = \eta_{\alpha} \sigma_{t\beta}^{\alpha} 
\sigma_{t\gamma}^{\beta} \xi^{\gamma}_{t} . 
\label{eq:6.19} 
\end{eqnarray} 
Therefore, putting the various ingredients together, we obtain: 
 
\begin{prop} 
The dynamics of the Hilbert space vector $\xi^{\alpha}_{t}$ 
that characterises the term structure in an admissible, 
arbitrage-free interest rate framework is governed by the 
stochastic differential equation 
\begin{eqnarray} 
{\rm d}\xi^{\alpha}_{t} = \left( {\tilde D}^{\alpha}_{\ \beta} 
- \octa \sigma^{\alpha}_{t\gamma}\cdot
\sigma^{\gamma}_{t\beta}\right)\xi^{\beta}_{t}{\rm d}t 
+ \half \sigma^{\alpha}_{t\beta}\xi^{\beta}_{t}\cdot 
\left( {\rm d}W_{t} + \lambda_{t}{\rm d}t\right)  , 
\label{eq:6.20} 
\end{eqnarray} 
where ${\tilde D}^{\alpha}_{\ \beta}$ is given as in 
{\rm (\ref{eq:6.15})}, and the adapted operator-valued 
process $\sigma_{t\alpha\beta}$ is expressible in the form 
\begin{eqnarray} 
\sigma_{t\alpha\beta} = \nu_{t\alpha\beta} - \left( 
\frac{\nu_{\gamma\delta}\xi^{\gamma}_{t}\xi_{t}^{\delta}}
{g_{\gamma\delta}\xi_{t}^{\gamma}\xi_{t}^{\delta}} \right) 
g_{\alpha\beta} ,  
\label{eq:6.21} 
\end{eqnarray} 
where $\nu_{t\alpha\beta}$ is an arbitrary adapted 
operator-valued process. 
\end{prop} 

This result shows that the evolution of the yield curve can be viewed 
consistently as a process on the positive orthant of the unit 
sphere in Hilbert space, and thus gives rise to an entirely new 
way of understanding the dynamics of the term structure. The 
purpose of the quadratic term in the drift of (\ref{eq:6.20}) is 
to keep the process on the sphere, and in the absence of the 
term involving the operator $D^{\alpha}_{\ \beta}$ we would have 
a general local martingale on the sphere ${\cal S}$ with respect to 
the risk-neutral measure, where the martingale property on 
${\cal S}$ is characterised in a standard way by use of the 
techniques of stochastic differential geometry (see, e.g., Emery 
1989, Ikeda $\&$ Watanabe 1989, Hughston 1996). The term involving 
the operator 
$D^{\alpha}_{\ \beta}$ splits into a 
symmetric and an antisymmetric part. The drift generated by 
the antisymmetric part of $D^{\alpha}_{\ \beta}$ is generated 
by a symmetry of the sphere ${\cal S}$. The drift generated 
by the symmetric part of $D^{\alpha}_{\ \beta}$, on the other 
hand, is a negative gradient vector field orthogonal to surfaces 
in ${\cal S}$ generated by level values of the short rate $r_{t}$. 
This term therefore creates a tendency for the vector $\xi^{\alpha}$ 
to drift towards a lower interest rate, a property of the negative 
gradient field which is then counterbalanced by the effects of the 
diffusive term.

\vspace{0.5cm} 

\begin{center} 
{\bf Acknowledgements} 
\end{center}
\vspace{0.2cm}

LPH acknowledges the hospitality of the Finance Department of 
the Graduate School of Business of the University of Texas at 
Austin, where part of this work was carried out. DCB gratefully 
acknowledges financial support from The Royal Society.

\end{document}